\documentclass{article}
\usepackage[utf8]{inputenc}
\usepackage{amsmath}
\usepackage{amssymb}
\usepackage{stmaryrd}
\usepackage{textcomp}
\usepackage{fancyvrb}
\usepackage{mathrsfs}
\usepackage{graphicx}
\usepackage{bm}
\usepackage{enumitem}
\usepackage{changepage}
\usepackage{mathtools}
\usepackage{setspace}
\usepackage[american]{babel}
\usepackage[style=apa]{biblatex}
\DeclareLanguageMapping{american}{american-apa}
\usepackage{csquotes}
\usepackage{indentfirst}
\usepackage{xypic}
\usepackage{istgame}
\usepackage{hyperref}
\usepackage{authblk}
\usepackage{geometry}
\renewbibmacro{in:}{}
\DeclareFieldFormat{pages}{#1}

\DeclareMathOperator*{\mini}{min}

\newtheorem{example}{Example}
\newtheorem{theorem}[example]{Theorem}
\newtheorem{proposition}[example]{Proposition}
\newtheorem{corollary}[example]{Corollary}

\newtheorem{remark}[example]{Remark}

\title{Preferences on Ranked-Choice Ballots}
\author{Brian Duricy\thanks{bduricy@alumni.cmu.edu}}
\date{December 19, 2022}

\begin{document}

\maketitle

\begin{abstract}
    This paper formalizes the lattice structure of the ballot voters cast in a ranked-choice election and the preferences that this structure induces. These preferences are shown to be counter to previous assumptions about the preferences of voters, which indicate that ranked-choice elections require different considerations for voters and candidates alike. While this model assumes that voters vote sincerely, the model of ranked-choice elections this paper presents allows for considerations of strategic voting in future work.
\end{abstract}

\hspace{3mm}\textbf{JEL Codes:} D71, D72

\hspace{3mm}\textbf{Keywords:} Ranked-choice voting, preferences, lattice theory

\section{Introduction}\label{intro}

Social choice models and results usually require strict preference relations, or those where every alternative is uniquely ranked with respect to the others. This includes the subset of social choice theory dedicated to voting, despite real-life elections that use ranked-choice voting\footnote{Elections between multiple candidates in which voters are required to vote for one and only one candidate satisfy this trivially.} either mandating non-strict preference relations or showing that voters effectively vote as if this is the case. Regarding the former, the 2021 Primary Elections for New York City Mayor allowed voters to rank up to five candidates (\textcite{NYC}), while the Democratic Primary had 13 total candidates to choose from (not including write-ins). Regarding the latter, \textcite{KGF} list 17 ranked-choice elections and, amongst them, the highest average percentage of candidates on a ballot who were ranked was slightly above $80\%$. Experimental results such as \textcite{Nielson} similarly show that respondents generally do not approach ranking all---or even most---of the candidates. Ranked-choice elections serve as a compelling counterexample against mandating strict preference relations in all social choice models. The preferences that do appear in ranked-choice elections are an example of the preferences studied in \textcite{Kreps}.

This paper focuses on the structure of the ballots used in these elections, referred to as \textit{ranked-choice ballots}. A ranked-choice ballot is the result of a voter having a \textit{top-truncated order} (or, alternatively, such as in \textcite{FL}, a \textit{top order}) over the set of candidates. These terms are fully defined in \hyperref[model]{Section 2}, but the intuition is that not all alternatives must be uniquely ranked. Whereas previous work on top-truncated preferences like \textcite{AAL} and \textcite{TE} have focused on scoring rules associated with these preferences, this paper examines the more foundational order-theoretic properties\footnote{A complete treatment of order, and more specifically lattice, theory can be found in \textcite{CML} and \textcite{Gratzer}, respectively.} that arise from equipping a set with a top-truncated order. As \textcite{TUK} prove that differing ballot lengths can produce different winners in the same instant-runoff election, determining which scoring rule to use is an important related line of research.

Another area of research on top-truncated preferences focuses on computational questions (e.g., \textcite{ML}). Top-truncated preferences also necessitate a discussion of results that do not require a lattice, as similar work like \textcite{CE2} is based upon a lattice rather than a semilattice. That a top-truncated set is a join semilattice is the paper's first result, and one that informs the rest of the paper's findings. The results in \hyperref[results]{Section 3} follow a unique and smooth path from lattice theory to utility functions, stopping along the way to provide novel applications of results from the preference and voting theory literature.

The pairing of lattice theory with preference relations is common, and this paper contributes to this literature by focusing on antitone preference relations. Ranked-choice voting motivates the need for an exploration into if---and how---results from this literature apply to a context that is suited for antitone preferences. This paper is the first to identify the connection between top-truncated preferences and ranked-choice voting, and it connects multiple strands of literature that have previously existed somewhat independently of one another. With an understanding of some mathematical properties of ranked-choice ballots and the preferences that define them, normative work regarding the value of ranked-choice voting vis-a-vis other voting methods will be enhanced.

\section{The Model and Additional Terminology}\label{model}

\subsection{The Model}\label{model2}

A \textit{ranked-choice election} $(V, C, \succsim)$ consists of a (possibly infinite) set of voters, $V$, a finite set of at least three candidates, $C$, and a complete top-truncated order profile for $(V, C)$, $\succsim$, which assigns to each voter $v \in V$ a complete top-truncated order on $C$\footnote{This is a specification of the \textcite{Osborne} construction, where a \textit{collective choice problem} $(N, X, \succsim)$ is defined with $N$ a set of individuals and $X$ a set of alternatives, with the latter two sets functioning analogously to the sets $V$ and $C$, respectively. $\succsim$ in \textcite{Osborne} is a preference profile, with each $\succsim$ a preference relation, defined to be a complete and transitive binary relation.}. We define $(C, \succsim_v)$ as a \textit{ranked-choice ballot} for voter $v$\footnote{In the \hyperref[results]{Results} section of the paper, the specification notation will be omitted, but the results apply to the individual ballot of each voter.}. In general, $(C, \succsim_v)$ is a \textit{ballot}, with the type of order profile $\succsim$ unspecified. Each voter can rank as many candidates (i.e., declare these candidates distinguishable to the others) as they wish, but they must rank at least one candidate. \hyperref[Example 1]{Example 1} below provides a sample ranked-choice ballot and its lattice representation.

\begin{example}\label{Example 1}
\normalfont Let $C = \{a, b, c, d, x, y, z\}$ and let voter $v \in V$'s preferences over $C$ be $x \succ y \succ z \succ a \sim b \sim c \sim d$. This is alternatively represented as $v$ ranking candidate $x$ first, $y$ second, $z$ third, and candidates $a$, $b$, $c$, and $d$ unranked and tied for fourth. The lattice construction of this ballot is shown below; straight lines indicate strict preference between candidates and wavy lines indicate indifference between candidates.
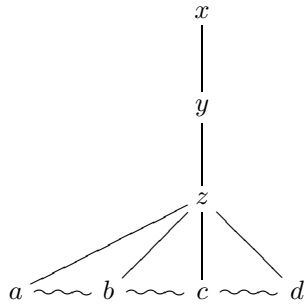
\begin{align*}
\xymatrix{
& & x \ar@{-}[d] & \\
& & y \ar@{-}[d] & \\
& & z \ar@{-}[d]\ar@{-}[dll]\ar@{-}[dr]\ar@{-}[dl] & \\
a \ar@{~}[r] & b \ar@{~}[r] & c\ar@{~}[r] & d
}
\end{align*}
It should be clear that these preferences exhibit the same ``desire for flexibility" (\textcite{Kreps}, p. 566) studied in Kreps' paper. If a set of candidates is a subset of another, the preferences on the latter are strictly preferred to the former if a ranked candidate is a member of the latter and not the former, and weakly preferred if the only members of the latter that are not members of the former are (additional) unranked candidates.
\end{example}

The utility function this paper uses assumes that voters do not vote strategically---i.e., a candidate is ranked above another candidate if and only if the voter would receive a greater utility from the former candidate winning the election than the latter; similarly, if multiple candidates are unranked, then each of those candidates winning provides the same utility to the voter. This can be formalized as follows: $u : C \shortrightarrow \mathbb{R}$ such that for all $x, y \in C$,

\begin{center}
    \begin{align*}
    u(x) > u(y) & \Leftrightarrow x \succ y \\
    u(x) = u(y) & \Leftrightarrow x \sim y
    \end{align*}
\end{center}

Additional terminology is needed for a full connection to the results of this paper and are defined in the \hyperref[terminology]{next subsection}.

\subsection{Additional Terminology}\label{terminology}

The concepts in this subsection can be divided into two parts: one that focuses on the order- and lattice-theoretic concepts needed, and one that focuses on the concepts regarding the utility function used in this paper.

A \textit{partial order} is a reflexive, transitive, and asymmetric binary relation. A set with a partial order is a \textit{partially ordered set}. A binary relation $\succsim$ is \textit{monotone} if for all $x, y \in C$, $x \geq y \Rightarrow x \succsim y$ and \textit{antitone} if for all $x, y \in C$, $x \geq y \Rightarrow x \precsim y$. If a candidate $x$ is \textit{preferred} to candidate $y$ by a voter $v$, we write $x \succ_v y$. If candidates $x$ and $y$ are \textit{indistinguishable} to voter $v$, then we write $x \sim_v y$. In the \hyperref[results]{results section} of this paper, it is sometimes easier to refer to candidates as \textit{ranked} or \textit{unranked}; the former refers to candidates that are not indistinguishable to any other candidate\footnote{Except, of course, itself, by the reflexivity of the binary relation.}, whereas the latter refers to candidates that are indistinguishable to at least one other candidate. A voter who ranks all candidates except for one trivially causes that last candidate to be ranked as well.

A \textit{weak order} is a partial order where indistinguishability is transitive. A \textit{top-truncated order} is a weak order where only the minimal elements are indistinguishable to one another, and a set with a top-truncated order is a \textit{top-truncated set} A partial order where every pair of elements are comparable is a \textit{complete partial order}. A partially ordered set where no pair of elements are indistinguishable is a \textit{totally ordered set}.

A \textit{join semilattice} is a partially ordered set where the least upper bound of each two elements in the set exists\footnote{I.e., for all $x, y \in C$, there exists $z$ such that $z = \sup\{x, y\} = x \vee y$.}. An element $x \in C$ is \textit{join-irreducible} if there exists a unique element $y \in C$ such that $x$ covers $y$\footnote{I.e., $x \succ y$ and there does not exist an element $z \in C$ such that $x \succ z \succ y$.}. Conversely, an element $x \in C$ is \textit{meet-irreducible} if there exists a unique element $y \in C$ such that $x$ is covered by $y$. An element is an \textit{atom} if it covers the least element of the set and is a \textit{co-atom} if it is covered by the greatest element of the set.

\begin{remark}\label{Remark 1}
\normalfont Some elementary lattice-theoretic properties of top-truncated sets are noted here without proof. If a top-truncated set has a join-irreducible element, that element is also an atom. A top-truncated set with a join-irreducible element is a totally ordered set. Every top-truncated set contains $n-1$ meet-irreducible elements and has $1 \leq m \leq n-1$ co-atoms. $\Box$
\end{remark}

A binary relation $\succsim$ is \textit{modular} (or \textit{strongly quasisubmodular}) if for all $x, y \in C$, $x \sim (x \vee y) \Rightarrow x \vee z \sim (x \vee y) \vee z$. A \textit{representation of $\succsim$} is a function $u : C \shortrightarrow \mathbb{R}$ such that for all $x, y \in C$ $x \succsim y \Rightarrow u(x) \geq u(y)$ and $x \succ y \Rightarrow u(x) > u(y)$. A representation $u : C \shortrightarrow \mathbb{R}$ is \textit{submodular} for a join semilattice if for all $x, y \in C$ such that there exists a greatest lower bound\footnote{I.e., for all $x, y \in C$, there exists $z$ such that $z = \inf\{x, y\} = x \wedge y$.}, $u(x \wedge y) + u(x \vee y) \leq u(x) + u(y)$. As some methods of ranked-choice voting are used to elect multiple candidates from a single election---with ``elect" here either meaning being one of the overall winners of the election or being one of the candidates who moves on to a head-to-head runoff---results relating to the representation of the preference relation are especially important for this context.

\textcite{Kalandrakis} focuses on a similar notion, rationalizability. $u : C \shortrightarrow \mathbb{R}$ is \textit{strictly rationalizable} if $u(x) > u(y)$ for each pair $x, y \in C$ such that $x \succsim y$ and \textit{rationalizable} if $u(x) \geq u(y)$ for each pair $x, y \in C$ such that $x \succsim y$. Clearly if a ballot has multiple unranked candidates, this leads to $u$ not being strictly rationalizable. $u : C \shortrightarrow \mathbb{R}$ is \textit{almost strictly rationalizable} if it is rationalizable over all pairs $x, y \in C$ and strictly rationalizable for each pair $x, y \in C$ such that it is not the case that $x \succsim y$ and $y \succsim x$. As should be clear from the definitions already provided, this allows for some results to be applied to the context of ranked-choice voting. Finally, $u : C \shortrightarrow \mathbb{R}$ is \textit{strictly concave} if $u(\lambda x + (1 - \lambda)y) > \lambda u(x) + (1 - \lambda)u(y)$ for all $x, y$ with $x \neq y$ and for all $\lambda \in (0, 1)$, and \textit{strictly quasiconcave} if $u_i(\lambda x + (1 - \lambda)y) > \mini\{u_i(x), u_i(y)\}$ for all $x, y$ with $x \neq y$ and for all $\lambda \in (0, 1)$.

While strategic voting is a usual feature of research regarding ranked-choice voting, providing a utility function that reflects this is beyond the scope of this paper. Contextualizing the results of this paper with a utility function for strategic voting is an area of future research. Strategic voting also potentially complicates analyses that rely on the preference relation being monotone, such as \textcite{CE} and \textcite{CMY}. The structure of a ranked-choice ballot for a voter who votes sincerely reflects an antitone preference relation---the candidate (say, $x$) that would provide the voter with the greatest utility is ranked $1$, descending until the candidate (or candidates) who would provide the voter with the least utility (say, without loss of generality, $y$) is ranked $k$; so $y > \mathellipsis > x \Rightarrow u(y) < \mathellipsis < u(x)$. \textcite{CVZ} provide a result that is dualized below that is similar to one found in \textcite{CE}, but for an antitone preference relation.

\section{Results}\label{results}

Having defined a ranked-choice ballot above, we begin providing results by formally connecting it to a well-known mathematical structure.

\begin{theorem}\label{Theorem 1}
If a ballot is a ranked-choice ballot, then it is a join semilattice.
\end{theorem}
\textbf{Proof:} Let $(C, \succsim)$ be a ranked-choice ballot. Since it is a top-truncated set, it necessarily is a partially ordered set. So all that remains to be shown is that the join exists for each pair of candidates. Let $x, y \in C$. If $x$ and $y$ are distinguishable, then, without loss of generality, say $x \succ y$; $x = x \vee y$ immediately follows. If $x$ and $y$ are not distinguishable, i.e., $x \sim y$, there exists at least one other candidate in the election, say, $z$, since at least one candidate must be ranked. If $z$ is the only other candidate, then $z \succ x \sim y$, which in turn means that $z \succ x$ and $z \succ y$. So $z = x \vee y$ similarly follows. If multiple candidates are ranked, for the previous relationships to hold, select $z$ as the candidate ranked last amongst them; $z = x \vee y$ again follows. Therefore, the join exists for each pair of candidates. Hence, $(C, \succsim)$ is a join semilattice. $\Box$

\vspace{2mm}

With a substantive literature on preferences over semilattices, this result is the first to highlight the connection to ranked-choice voting. This, along with a couple of other features inherent in ranked-choice ballots, unlocks some important properties of the utility function associated with these ballots. These properties support the usage of ranked-choice voting as a way to increase the overall utility from an election. The \hyperref[Proposition 1]{next result} is the second of the features needed to satisfy the conditions for the first result regarding utility functions.

\begin{proposition}\label{Proposition 1}
If $\succsim$ is a top-truncated order, then it is modular.
\end{proposition}
\textbf{Proof:} Let $\succsim$ be a top-truncated order on $C$ and let $x, y \in C$ such that $x \sim (x \vee y)$. Then, since $x \vee y$ must be a ranked candidate and $x \sim (x \vee y)$, $x$ must be $x \vee y$ since a ranked candidate can only be indistinguishable to itself. So, since $x = (x \vee y)$, $x \vee z \sim (x \vee y) \vee z$, satisfying the definition of modularity. Therefore, top-truncated orders are modular. $\Box$

\vspace{2mm}

Ranked-choice ballots are proven to be (finite) join semilattices, with top-truncated orders being modular (or strongly quasisubmodular). Additionally, the top-truncated orders in this model are complete, and thus a type of complete preorder. Finally, as the preferences in this paper are antitone, they are (weakly) decreasing. Therefore, ranked-choice ballots have all of the necessary conditions to satisfy the \hyperref[Proposition 2]{following proposition}.

\begin{proposition}\label{Proposition 2}
(Dual of \textbf{Corollary 2} from \textcite{CVZ}.) For a complete preorder $\succsim$ on a finite join semilattice $(C, \succsim)$, the following are equivalent: 
\begin{enumerate}
    \item $\succsim$ is weakly decreasing and strongly quasisubmodular.
    \item $\succsim$ has a weakly decreasing and submodular representation. $\Box$
\end{enumerate}
\end{proposition}

We can then establish the \hyperref[Corollary 1]{subsequent corollary}.

\begin{corollary}\label{Corollary 1}
A ranked-choice ballot has a submodular representation. $\Box$
\end{corollary}

We next show that the preferences over ranked-choice ballots allow for a result from \textcite{Kalandrakis} to hold that further characterizes the utility function associated with these ballots. It helps to first define the following concepts: let $P \subseteq C \times C$ be the set of pairs of candidates, with $(x, y) \in P$ meaning that $x$ is (weakly) preferred to $y$. The potential weakness of preferences is necessary, as indistinguishable candidates $x, y$ are in $P$ as the separate pairs $(x, y) \in P$ and $(y, x) \in P$. Let $Y(P)$ be the set of candidates that are (weakly) preferred to at least one other candidate, and $N(P)$ be the set of candidates that are (weakly) not preferred to at least one other candidate. Finally, let $\mathcal{E}(C)$ be the set of extreme points, or the candidates that are unable to be written as a strict convex combination of candidates in $C$; $\mathcal{E}(C(P))$ indicates that these candidates are part of at least one pair in $P$. The \hyperref[Theorem 3]{following theorem} is needed to apply the remainder of the result from \textcite{Kalandrakis}. A necessary fact about the set of extreme points regarding ranked-choice ballots is that the highest-ranked candidate in a set and the lowest-ranked candidate are the extreme points; if a set has multiple unranked (i.e., lowest-ranked) candidates, each of those candidates are in the set of extreme points, unless the set contains only unranked candidates, as that set would then have no extreme points.

\vspace{6mm}

\begin{theorem}\label{Theorem 3}
For all nonempty $P' \subseteq P$, either 
\begin{enumerate}
    \item there exists $x \in \mathcal{E}(C(P'))$ such that $x \not \in Y(P')$
    \item there exists a nonempty $P'' \subseteq P'$ such that $N(P'') = Y(P'') \subseteq \mathcal{E}(C(P'))$ and $Y(P'') \cap Y(P' \setminus P'') = \varnothing$.
\end{enumerate}
\end{theorem}
\textbf{Proof:} The proof proceeds in three parts which correspond to the three possible combinations of candidates in $P'$---all ranked, at least one ranked and at least one unranked, and no ranked candidates. First, let $P' \subseteq P$ such that all candidates in $P'$ are ranked. Then, $\mathcal{E}(C(P')) = \{x, y\}$ with $x$ the highest-ranked and $y$ the lowest-ranked candidate; $Y(P') = P' \setminus \{y\}$; and $N(P') = P' \setminus \{x\}$. Clearly, as $y \in \mathcal{E}(C(P'))$ but $y \not \in Y(P')$, the conditions hold. Next, let $P' \subseteq P$ consist of at least one ranked candidate, $x$, and at least one unranked candidate, $y$. Then, without loss of generality, $\mathcal{E}(C(P')) = \{x, y\}$; $Y(P') = P' \setminus \{y\}$; and $N(P') = P' \setminus \{x\}$. Again, as $y \in \mathcal{E}(C(P'))$ but $y \not \in Y(P')$, the conditions hold. Finally, let $P' \subseteq P$ such that $P'$ consists of only unranked candidates. Then, $\mathcal{E}(C(P')) = \varnothing$. Similarly, $Y(P') = N(P') = P'$. So for any subset of $P'$, say, $P'' \subseteq P'$, also has $Y(P'') = N(P'').$ However, since all nonempty $P''$ are such that $N(P'') = Y(P'')$ and $N(P'') = Y(P'') \not \subseteq \mathcal{E}(C(P')) = \varnothing$, this satisfies the contrapositive of the second condition. $\Box$

\vspace{2mm}

The following result from \textcite{Kalandrakis} proves that ranked-choice ballots lead to voters having concave utility functions. As work on the strategic voting of candidates such as \textcite{Tajika} assumes that voters have convex utility functions, candidates as well as voters have an incentive to act differently in a ranked-choice election than in a traditional first-past-the-post election.

\begin{theorem}\label{Theorem 4}
(\textbf{Theorem 2} from \textcite{Kalandrakis}.) Let $C$ be the set of candidates and $P \subseteq C \times C$ be the voting record for a given voter. Then the following conditions are equivalent:
\begin{enumerate}
    \item For all nonempty $P' \subseteq P$, either there exists $x \in \mathcal{E}(C(P'))$ such that $x \not \in Y(P')$ or there exists a nonempty $P'' \subseteq P'$ such that $N(P'') = Y(P'') \subseteq \mathcal{E}(C(P'))$ and $Y(P'') \cap Y(P' \setminus P'') = \varnothing$.
    \item There exists a strictly concave utility function that almost strictly rationalizes $P$.
    \item There exists a strictly quasiconcave utility function that almost strictly rationalizes $P$.
    \item There exists a strictly concave utility function that rationalizes $P$.
    \item There exists a strictly quasiconcave utility function that rationalizes $P$. $\Box$
\end{enumerate}
\end{theorem}

\section{Conclusion}\label{conclusion}

This paper was the first to formalize the preferences of ranked-choice voting and explore what structure a ballot having those preferences takes. Top-truncated preferences elicit specific types of representations and utility functions; now that these have been identified, a more substantive appraisal of ranked-choice voting's value can be done. 

There are also multiple areas of future research that can build upon the results from this paper. \textcite{AAL} mention the need for normative work on top-truncated preferences, which is especially important because these preferences have been shown to be concave (and quasiconcave)---types of preferences not always assumed to reflect voters' actual preferences. Whether these preference types are affected if the utility function accounts for strategic voting is a valuable question to explore. \textcite{Coughlin} addresses utility functions for strategic voting, but in the context of candidates' utility functions rather than voters' utility functions. Ranked-choice voting provides both the opportunity for a voter to express their full set of preferences and the opportunity to vote strategically. This paper has explored theoretical properties associated with the former; the next step is to see if and where there is an intersection with the latter.

\newpage

\printbibliography

\end{document}